\documentclass{emulateapj}

\usepackage{epsfig}
\newcommand{\bma}{\begin{math}}
\newcommand{\ema}{\end{math}}
\newcommand{\beq}{\begin{equation}}
\newcommand{\eeq}{\end{equation}}
\newcommand{\bc}{\begin{center}}
\newcommand{\ec}{\end{center}}
\newcommand{\bit}{\begin{itemize}}
\newcommand{\eit}{\end{itemize}}

\def\n{{\bf  \hat n}}

\newcommand{\erfc}{{\rm erfc}}

\begin{document}

\title{The influence of non-uniform reionization on the CMB}

\author{Oliver Zahn\altaffilmark{1,3}, Matias
    Zaldarriaga\altaffilmark{1,2}, Lars Hernquist\altaffilmark{1} \&
    Matthew McQuinn\altaffilmark{1}}

\altaffiltext{1} {Harvard-Smithsonian Center for Astrophysics, 60 Garden
  Street, Cambridge, MA 02138, USA}
\altaffiltext{2} {Jefferson Physical Laboratory, Harvard University, 
  Cambridge, MA 02138, USA}
\altaffiltext{3} {e-mail address: ozahn@cfa.harvard.edu}

\begin{abstract}

We investigate the impact of spatial variations in the ionized
fraction during reionization on temperature anisotropies in the CMB.
We combine simulations of large scale structure to describe the
underlying density field with an analytic model based on extended
Press-Schechter theory to track the reionization process.  We find
that the power spectrum of the induced CMB anisotropies depends
sensitively on the character of the reionization epoch. Models that
differ in the extent of the ``patchy phase" could be distinguished by
future experiments such as the Atacama Cosmology Telescope (ACT) and
the South Pole Telescope (SPT).  In our models, the patchy signal
peaks at $l\simeq 2000$, where it can be four times larger than the
kinetic Sunyaev-Zel'dovich (kSZ)/Ostriker-Vishniac (OV) signal
($\Delta T_{\rm tot} \simeq 2.6 \mu K$).  On scales beyond $l \simeq
4000$ the total Doppler signal is dominated by kSZ/OV, but the patchy
signal can contribute up to 30\% to the power spectrum. The effect of patchy
reionization is largest on scales where the primordial CMB
anisotropies dominate. Ignoring this contribution could lead to
significant biases in the determination of cosmological parameters
derived from CMB temperature measurements. Improvements in the
theoretical modeling of the reionization epoch will become
increasingly important to interpret the results of upcoming
experiments.

\end{abstract}

\keywords{cosmology: theory -- cosmic microwave background -- large-scale structure}

\section{Introduction}
\label{intro}

Cosmic Microwave Background (CMB) anisotropy experiments have now
constrained temperature fluctuations down to scales of $4'$
\citep{Readhead:2004gy,Holzapfel:1999ee} resulting in a much improved
understanding of inhomogeneities during the time of decoupling at
$z\simeq 1100$ and of the global properties of the Universe. In
combination with data from Supernovae Ia, measurements of the
local expansion rate, galaxy clustering studies, and observations of
the Lyman $\alpha$ forest, the CMB data has lead to the establishment
of a ``standard" cosmological model. We live in a Universe dominated
by dark energy and with significantly more dark matter than
baryons. Structure grew from a scale independent spectrum of
primordial Gaussian fluctuations, in good agreement with predictions
of inflationary models.

In the CMB community, theoretical and experimental interest is
shifting to the study of secondary anisotropies on smaller angular
scales. These are caused by fluctuations in the distribution of
baryons and dark matter in the redshift regime $z \simeq 0-30$. Future
data on the different secondary effects will constrain the way
structure formation proceeded from the linear into the non-linear
regimes. Secondary anisotropies can be divided into three categories:
gravitational lensing effects, inverse Compton scattering of CMB
photons by hot plasma, and Doppler related anisotropies.

Mass concentrations along the line of sight such as clusters, sheets,
and filaments lead to weak deflection of the CMB photons, or
\emph{Gravitational Lensing} (for a review see chapter 9 of
\citealt{Bartelmann:1999yn}).  Gravitational lensing of the CMB can be
used as a cosmological probe in various ways. The lensing effect on
the CMB power spectrum could help break degeneracies between
cosmological parameters \citep{Metcalf:1997ad,Stompor:1998zj}. Lensing
induces departures from Gaussianity on CMB maps that can be used to
reconstruct the spatial distribution of the lensing mass
\citep{Zaldarriaga:1998te,Hu:2001kj}.  The CMB lensing signal can be
correlated with galaxy lensing shear providing additional
information. Thus, lensing can be used to probe the evolution of
gravitational clustering, constrain the properties of the dark energy,
the shape of the matter power spectrum, and neutrino physics (see
e.g. \citealt{vanWaerbeke:1999jd,Hu:2001fb,Kaplinghat:2003bh}).
  
\emph{Inverse Compton scattering} in the hot
  intracluster medium, also called the thermal Sunyaev-Zel'dovich
  effect \citep{Zeldovich:1969ff}, changes the spectrum of the CMB
  photons, leading to cold spots (decrements) in the microwave
  background at frequencies below $~217$ GHz and to hot spots
  (increments) at frequencies above. The effect is proportional to the
  line-of-sight integrated pressure. Because it is independent of
  redshift, it is a unique probe of collapsing structures out to
  $z\simeq 3$ (see Figure \ref{y_zdepend}). The thermal SZ effect has been
  measured in follow-up observations of X-ray clusters (see for
  instance
\citealt{Birkinshaw:1998qp,Carlstrom:1999hs}). 
Comparison of the cluster SZ
temperatures with the X-ray measured temperatures leads to constraints
on the angular diameter distance.
The SZ effect also leaves a signature in the CMB power spectrum on
small scales where the primordial CMB vanishes. BIMA (the Berkeley
Illinois Maryland Association) and CBI (the Cosmic Background Imager)
claim detections of this effect at $\geq 2 \sigma$ significance levels
\citep{Readhead:2004gy,Holzapfel:1999ee}. CBI infers a LSS clustering
amplitude on 8 Mpc/h scales of $\sigma_8=0.9$, which is at the upper
level of constraints from cluster observations and weak lensing (for a
recent compilation of experimental results see table 5 of
\citealt{Tegmark:2003ud}).  Future large angular scale SZ surveys such
as SPT promise to measure the cluster abundance as a function of mass
and redshift, which will offer the possibility of constraining the
matter density and the equation of state of the dark energy.

Finally, \emph{Doppler} related effects are produced by the scattering
of CMB photons off electrons moving as a result of the structure
formation process.  This process, also known as the kinetic
Sunyaev-Zel'dovich effect when applied to clusters of galaxies
\citep{Sunyaev:1980nv}, leads to hot or cold spots, depending on
whether the ionized baryons move toward or away from the observer. The
frequency dependence of the photons is left unchanged, except for tiny
relativistic corrections.  These ``Doppler" induced anisotropies are
the only known way to measure the high redshift large scale velocity
field.

Although our simulations account for each of these effects, our focus
in this paper will be on the part of the Doppler effect generated
during the epoch of reionization.  This epoch is currently not well
understood observationally.


The first year release of the WMAP data hinted at a large optical
depth owing to reionization, $\tau_{ri} =0.17 \pm 0.05$
\citep{Spergel:2003cb}, through measurement of the ``reionization
bump'' \citep{Zaldarriaga:1996ke} in the temperature polarization
cross correlation \citep{Kogut:2003et}.  WMAP places only a weak
constraint on the relative height of the acoustic oscillation peaks,
so these data alone are not sufficient for constraining $\tau_{ri}$,
because it is strongly degenerate with the tilt of scalar
fluctuations $n_s$ and the baryon density parameter $\omega_b=\Omega_b
h^2$.  In combination with smaller scale CMB measurements, the data
can be described by homogeneous reionization of the universe taking
place at redshift $z \simeq 14 \pm 3$, or a Thomson scattering optical
depth $\tau_{ri} \simeq 0.13 \pm 0.02$ (see
e.g. \citealt{Readhead:2004gy,Bond:2003ur}). This estimate is also robust
to the addition of other datasets to the analysis, such as the SDSS
galaxy clustering survey
\citep{Tegmark:2003ud} ($\tau_{ri} =0.124_{-0.057}^{+0.083}$) or the
Lyman alpha forest \citep{Seljak:2004xh} ($\tau_{ri} = 
0.133_{-0.045}^{+0.052}$). 

Since the large scale polarization measurements are sensitive only to
the line-of-sight integrated free electron density, they do not
require the time evolution of the ionized fraction to be a simple step
function.  Observations of the evolution of the Ly$\alpha$ optical
depth in the absorption spectra of the highest redshift quasars at $z
\simeq 6$ when combined with the CMB constraints favor a more
complicated ionization history. The observations show a distinct 
Gunn-Peterson trough \citep{Gunn:1965hd} and
point to a rapidly evolving neutral fraction indicative of the final
stages of reionization
\citep{Becker:2001ee,Fan:2001ff,White:2003sc,Fan:2004bn}.  Recently,
there has also been an interpretation of the relatively high
temperature of the Ly$\alpha$ forest at $z\simeq 2-4$ as evidence of
an order unity change in the ionized fraction at $z<10$
\citep{Theuns:2002yc,Hui:2003hn}, although this depends
on the properties of He II reionization \citep{Sokasian:2002jw}. 


Numerical simulations of the star formation epoch have shown that a
first generation of low metallicity (Population III) sources can
survive negative feedback from the UV background they produce if
shielded in massive halos for some time
\citep{Schaerer:2002yr,Bromm:2001ag}. If they are heavy, these stars could
produce an order of magnitude more ionizing photons per baryons than
``normal'' stars.  Besides a more complex temporal behavior, the
ionization fraction could have spatial variations; it could be
``patchy" for a period of time while reionization is not complete.
Naturally, we would like to have an observable that sheds light on the
duration and structure of the reionization epoch.


The shape of the large scale polarization signal in the CMB could be
used to constrain the evolution of the \emph{global} ionization
fraction, without resorting to information from secondary anisotropies
\citep{Zaldarriaga:1996ke,Hu:2003gh}.  However, these scales are
affected by sample variance limitations, so the available information
is rather limited. Moreover, before such an approach can be
successful, issues of foreground removal need to be clarified
\citep{Bennett:2003ca,deOliveira-Costa:2003pu}.  In any event, large
scale polarization measurements will not yield information about the
morphology of reionization, as this is confined to sub-degree scales.


Several analytical models for patchy reionization have been presented
in the literature.  They vary significantly in the assumed size and
time evolution of spherically shaped bubbles, and whether the ionized
regions are correlated
\citep{Knox:1998fp,Santos:2003jb} or not \citep{Gruzinov:1998un}.
Some models employ a prescription for correlating the ionizing sources by
using the bias that can be calculated for the dark matter halos in which
they presumably reside
\citep{Santos:2003jb}.
All authors agree that on scales below $4'$, where the primordial CMB signal
falls rapidly owing to photon diffusion,  the Doppler effect induced by
patchiness could contribute enough to use it as a tool to study the 
reionization epoch.


Modeling patchy reionization more accurately requires the use of
numerical simulations to incorporate effects of non-linear clustering
and radiative transfer. The morphology of the ionized regions depends
sensitively on the star formation model (see e.g. the reionization
study companion by \citealt{Barkana:2000fd}), source properties,
feedback processes, and radiative transfer (these effects have for
example been modeled in
\citealt{Abel:2001qs,Gnedin:2000gr,Ciardi:2003ia,Sokasian:2003au,
Sokasian:2004au}).
These simulations have given us evidence that reionization was much
less homogeneous than previously thought
\citep{Miralda-Escude:1998qs,Barkana:2000fd}. Reionization tends to
proceed from high to low density regions
\citep{Sokasian:2003au,Sokasian:2004au,Ciardi:2003ia} and 
recombinations seem to play
a role subdominant to large scale bias. Str\"omgren spheres of
neighboring protogalaxies overlap, and overdensities will harbor large
ionized regions. Because HII regions extend to larger radii than the
correlation length of galaxies, building an analytic model solely on
local galaxy properties appears to be difficult.  Besides
uncertainties over the physics in current radiative transfer
calculations of reionization, memory and CPU requirements pose a
serious limitation, so that these simulations have so far only been
performed as a post-processing step on scales of up to 10 Mpc/h,
corresponding to $\simeq 6$ arcminutes on the sky
\citep{Furlanetto:2003nf}. If the HII regions are of comparable size,
sample variance becomes a problem for these simulations.


In this paper, we use a hybrid approach between an analytic model for
the formation of HII regions, and smoothed particle hydrodynamics (SPH)
simulations of large scale structure.  The advantage over analytic
models of patchy reionization is that our sources follow the complex
clustering behavior of dark matter and baryons.  An advantage over
currently feasible full radiative transfer calculations of the
reionization epoch is that we can investigate the morphological
properties of reionization on scales an order of magnitude larger with
rather small memory and CPU requirements.  Our prescription does not
directly address many of the (uncertain) physical details of
reionization related to star formation, feedback processes,
clumpiness, recombinations, and radiative transfer. However, by
combining these properties into a single parameter, we are able to
explore the basic parameter dependencies, and we believe it is a good
starting point to help us understand the morphological properties of
cosmological reionization and its influence on the CMB.


We will describe the analytic model for the morphology of patchy
reionization suggested by
\citet{Furlanetto:2004nh} in the next section. 
In this description, the size distribution of HII regions is derived
in a way analogous to the Press-Schechter formalism for the halo mass
function, so that sources are biased towards large scale matter
concentrations.  The model has only one free parameter, the efficiency
of the first sources to ionize the surrounding medium.


In Section \ref{simul} we describe our implementation of the model
into SPH simulations as a post-processing step.  Since the costs in
memory and CPU consumption in this scheme are relatively low, one can
imagine testing a wide array of behaviors for the ionization efficiency
of the first sources against experiments.


One of the parameterizations we employ, in Section \ref{specific}, attempts to do justice to the
apparently contradicting measurements of a large integrated Thompson
optical depth in CMB polarization measurements that indicates an onset
of reionization at redshifts beyond $z=15$, and a rapid decline of the
neutral fraction around $z=6-7$, seen in measurements of the
Gunn-Peterson optical depth by measuring spectra of distant
quasars. We model this by assuming two succeeding reionization epochs,
in which HII regions first are produced by low metallicity (Pop. III)
stars, and later expand further during an epoch dominated by Pop. II stars, so
that on the whole patchiness lasts longer.


In Section \ref{exper} we investigate whether the next generation
of ground based bolometric arrays will be able to measure reionization
and distinguish between different scenarios. The specifications of the
Atacama Cosmology Telescope (ACT) \footnote{see http://www.hep.upenn.edu/~angelica/act/act.html} \citep{Kosowsky:2004sw} and the
South Pole Telescope (SPT)\footnote{see
http://astro.uchicago.edu/spt/} \citep{Ruhl:2004kv} will be used to predict how well these
experiments can measure CMB temperature power spectra in the region
where secondary effects dominate over primordial CMB anisotropies.  We
find that ACT/SPT should be able to distinguish between sudden
homogeneous reionization and patchy reionization at a
high level of significance,
even if the total optical depth in both scenarios were the same.


Since the patchy signal peaks on scales where the primordial
anisotropy dominates, cosmological parameter estimation may become
biased. We will show in Section \ref{fisher} that precision
experiments such as Planck should take this bias into account either
by focusing on their polarization data or by adding a reionization
parameter to their analysis.


Our numerical simulations of large scale structure and the power
spectra we used to generate realizations of the primordial CMB assume
a cosmology in agreement with WMAP constraints: $\Omega_{dm}=0.26$,
$\Omega_\Lambda=0.7$, $\Omega_{b}=0.04$, fluctuations on 8Mpc/h scales
of $\sigma_8=0.9$, and no tilt ($n_S=1$) or running ($\alpha_S=0$)
of the spectral index.

\section{Analytic model for patchy reionization}
\label{model}

We review the model for the growth of HII regions during reionization
proposed by \citet{Furlanetto:2004nh}.  The standard way to describe
ionization of the IGM is to associate an HII region with a single
galaxy. The size distribution of ionized regions then follows directly
from the halo mass function when the ansatz $m_{ion}=\zeta m_{gal}$ is
made, where $m_{gal}$ is the mass in a collapsed object. The parameter
$\zeta$ is the efficiency factor for ionization, for example composed
as $\zeta=f_{esc}f_* N_{\gamma/b} n_{rec}^{-1}$, where $f_{esc}$ is
the escape fraction of ionizing photons from the object, $f_*$ is the
star formation efficiency, $N_{\gamma/b}$ the number of ionizing
photons produced per baryon by stars, and $n_{rec}$ is the typical
number of times a hydrogen atom has recombined. The efficiency factor
is a rough combination of uncertain source properties, but
encapsulating a variety of reionization scenarios in it can be regarded
as a starting point to gain insights in the morphological properties
of the partly ionized phase.

According to the extended Press-Schechter model \citep{Bond:1990iw},
the collapsed fraction (or the fraction of baryons that lie in
galaxies) in a region of size r depends on the mean overdensity of
that region, $\overline{\delta}_r$, as
\beq
f_{coll}(m_{\rm min})=\erfc\left[\frac{\delta_c(z)-\overline{\delta}_r}{\sqrt{2[\sigma^2(r_{min})-\sigma^2(r)]}}\right]
\, .
\label{collfract} \eeq
Here, $\sigma^2(r)$ is the linear theory rms fluctuation on scale r and 
${\rm{r}_{min}}$ is taken to be the radius that encloses the mass
${\rm{m}_{min}}$ (at average density $\overline{\rho}$) corresponding to
a virial temperature of $10^4$K, at which atomic hydrogen line cooling
becomes efficient and $\delta_c(z)$ is the numerical factor $1.686$
scaled to today using linear theory. The redshift dependence of the
minimum mass (see \citealt{Barkana:2000fd}) can be described by
\beq
m_{\rm min} \simeq 3.0 \times 10^9 (1+z)^{-1.5} M_\odot \, .
\eeq

The mass in galaxies that can create enough luminous sources to fully
ionize all hydrogen atoms is inversely proportional to the ionizing
efficiency, so one requires that
\beq
f_{coll} \geq \zeta^{-1} \, .
\eeq
Hence, we can define a barrier which fluctuations have to cross for
their baryonic content to become ionized:
\beq
\delta_r \geq \delta_x(m,z) \equiv \delta_c(z)-\sqrt{2}
\erfc^{-1}(\zeta^{-1}) [\sigma^2(r_{\rm min})-\sigma^2(r)]^{1/2} \, .
\label{barrier} 
\eeq
Because it assumes Gaussian fluctuations on the mass scale $m$, this
formalism can be applied only to mass scales larger than the typical
size of collapsed objects.

\section{Simulation of Patchy Reionization}
\label{simul}


Our approach is to use large scale simulations of the cosmic web as a
basis for applying the model described in the last section as a
post-processing step to generate HII regions during the reionization
phase.  The underlying large scale structure was simulated using the
parallel Tree-PM/SPH solver GADGET \citep{syw01}, based on the fully conservative implementation of
SPH by \citet{sh02}. 
Here, we used the results of runs with
a boxsize of $100 {\rm Mpc}/h$ and $216^3$ particles, with
parameters corresponding to the G-series runs of \citet{sh03}.  In what follows, we employ a simulation that
included only ``adiabatic'' gas physics; i.e. the gas can heat
or cool adiabatically and be shocked, but we do not include
radiative effects or the consequences of star formation
and associated feedback processes.
Snapshots of the
simulation are produced every light crossing time interval. This leads
to 77 outputs between $z=0$ and $z=20$. These simulations are
described in more detail in \citet{White:2002wp}.


To generate inhomogeneously ionized regions according to the model
described in Section \ref{model}, we discretize the matter
fluctuations in our simulation boxes into $256^3$ cells. The
overdensity, $\delta$, is smoothed with a top hat window function.

To find the smoothing scale at which a given cell is ionized we have
to keep in mind that it could also be ionized by photons originating
from neighboring regions. Thus we should smooth the density on all
possible scales to see whether a given point was above the ionization
threshold of equation (\ref{barrier}) for some smoothing scale.  In
practice, we start at large radii (comparable to the simulation box
size) and record the smoothed overdensity as we smooth logarithmically
on progressively smaller scales. When a cell first crosses the barrier
(which depends only on redshift and ionization efficiency), it is
deemed ionized. If later, as a function of decreasing scale, it
crosses the barrier downwards, this means that the region had been
initially ionized by a neighboring overdensity.

\begin{figure}
\bc
\epsfig{file=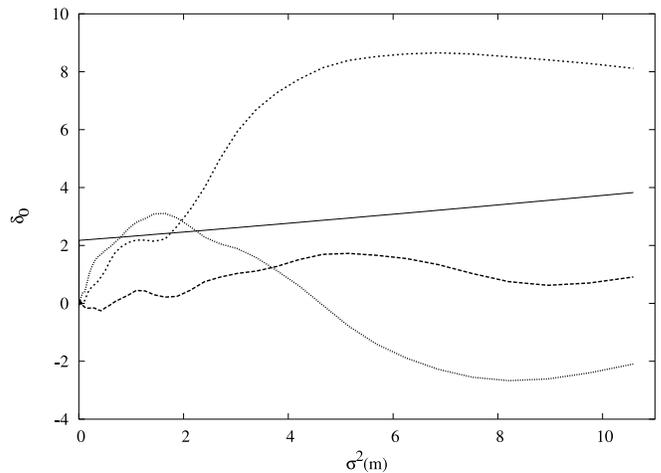,width=7cm,angle=-90}
\ec
\caption{The behavior of the overdensity scaled to today, $\delta_0$,
for four regions from our $256^3$ calculation. The abscissa in this
plot is the rms fluctuation, dependent on the applied smoothing
scale. The solid curve is the barrier given from
equation (\ref{barrier}).
The region corresponding to the short dashed curve crosses the
barrier for ionization at $\sigma^2(m) \simeq 1.8$. The region
described by the dotted curve wanders below the barrier again, but has
been ionized by sources in a neighboring region with higher
density. Finally the long dashed curve represents a region in which
the gas stays neutral at this particular redshift
($\overline{z}=14.2$).}
\label{walks}
\end{figure}

Figure \ref{walks} shows the random walk of $\delta_0$, the density
fluctuation scaled to the present using linear theory, for three different
regions inside our box (at redshift $z=14.2$) with increasing
$\sigma^2(m)$ (decreasing smoothing radius $r$). It also shows the
ionization efficiency dependent barrier, from equation (\ref{barrier}),
as the solid curve. The short-dashed curve describes a region of high
overdensity that self-ionizes. In the dotted curve at $\sigma^2(m)
\simeq 2.2$ the barrier is crossed downwards, so the volume element was
ionized by sources in neighboring cells. The long-dashed curve
corresponds to an element that did not ionize at this redshift.

\begin{figure*}
\bc
\epsfig{file=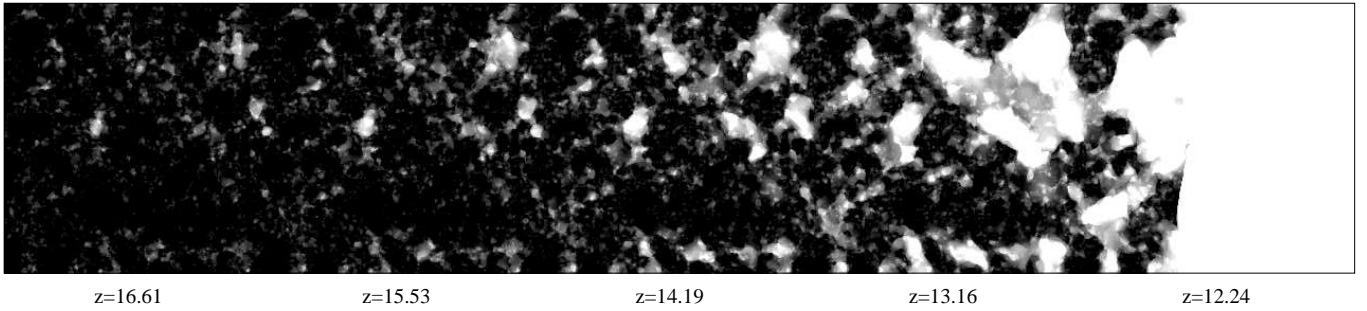,width=18cm}
\ec
\caption{Time development between redshifts $z=12$ and
  $z=17$ of an 8Mpc/h deep slice through an inhomogeneously
  ionized $\zeta=60$ (constant) box, with increasing ionization
  fraction. The outputs with periodic boundary conditions are simply
  stacked behind each other here. In order to obtain realistic maps,
  the individual outputs where randomly rotated and translated in our
  simulations. At the low redshift end, the bubbles become comparable
  in size to the box.}
\label{movie}
\end{figure*}

To sample a wide range of bubble sizes, and obtain a smoothly varying
ionization fraction, we varied the barrier in radial direction
throughout each box by adjusting the redshift dependent minimal
ionization radius.  The reionization information for each of the
$256^3$ cells is stored. A $8 \, {\rm Mpc/h}$ deep cut through the
stacked outputs between $z \simeq 12$ and $z \simeq 17$ is shown in Figure
\ref{movie}, for one of the models we will describe in the next
section, in which the ionization efficiency is constant,
$\zeta=60$. The periodic boundary conditions of our simulation boxes
are apparent in this plot.

To compute the electron scattering and gravitational lensing effects from the content of the simulation boxes, we only need to store two-dimensional maps of the product of pressure and volume (Compton scattering), the velocity weighted free electron density (Doppler effect), and the matter density (lensing). We simulate regions on the sky with an angular extent of $1^\circ$ each, at a resolution of $256^2$ pixels. In the small angle approximation we can simply project the content of each simulation box onto a plane in its center. At high redshifts, $z>5.7$, the field of view exceeds the boxsize, so we use the periodic boundary conditions to cover it. During the patchy regime, only those gas particles that are located in an ionized cell according to the information we stored earlier contribute to the signal. At later times all particles are assumed ionized. 
We translate and rotate the positions and velocities of the particles randomly before doing the projection onto the plane. The gas particle properties are distributed over surrounding pixels weighted by the SPH kernel. For dark matter particle masses we use the cloud-in-cell algorithm. Photons are traced through the planes from a regular grid close to the observer towards the last scattering surface. We produced Gaussian random fields of the primordial CMB using spherical harmonic decomposition coefficients generated with the publicly available code CMBFAST \citep{Seljak:1996is} \footnote{http://www.cmbfast.org}.


In models of patchy reionization, CMB anisotropies are caused by
two types of contributions induced by Doppler scatterings: perturbations in the
baryon density $\omega_{\rm b}$, given by $\delta \omega_{\rm b}
=\rho_{\rm b}/\rho_{\rm tot}-1$, and local changes in the ionization
fraction $\delta x_e$.  These produce a change in temperature of the
CMB blackbody.  The total contribution to the temperature anisotropy
is given by the integral over conformal time 
\footnote{During
homogeneous reionization and at lower redshifts, where galaxy clusters
are present, this is traditionally called the kinetic
Sunyaev-Zel'dovich effect, at higher redshifts the Ostriker-Vishniac
\citep{OV} effect}
\beq
\frac{\Delta T_{kSZ}}{T_{CMB}} (\n) = - \frac{\sigma_T}{c} \overline{n}_p (\eta_0) \int d
\eta [a^{-2} e^{-\tau_{ri}(\eta)} \bar x_e(\eta)] \n \cdot q 
\eeq
with
\beq
q=(1+\frac{\delta x_e}{\bar x_e(\eta)}) (1+\delta \omega_{\rm b}) \mathbf{v}
\, .
\eeq
The Thomson scattering optical depth is $\sigma_T$, $\tau_{ri}$ is the optical depth from the observer to conformal time $\eta$, and $\n$ is the line of sight unit vector.


To improve our understanding of reionization, various
parameterizations of $\zeta$ and its derivatives may be compared to
data from the next generation of experiments.  For each redshift, the
only free parameter in our model of patchy reionization is the
efficiency of the first sources which ionize the surrounding
intergalactic medium.  In contrast to radiative transfer simulations,
our implementation is not limited by CPU or memory related
problems. For each box we need to Fourier transform $\simeq 50$ times
back and forward to do the smoothing. This takes $\simeq 100$ minutes
per model on a Xeon 3.2 GHz computer. The implementation may be
parallelized easily, hence the realization of a large number of
parameterizations of $\zeta$ is feasible.


Even in the context of our simplified model, there are two additional
limitations to the accuracy of our present simulations. First, because
the mean overdensity on scales larger than the simulation box is
artificially set to zero in numerical simulations, in order to achieve
periodic boundary conditions, we expect a systematic bias in the
overall ionization fraction. \citet{Barkana:2003qk} estimated this
effect and showed that it should be less than 1\% in simulations of
100 {\rm Mpc}/h size. A different bias arises from our finite mass
resolution. Our lack of structure on scales smaller than
$L_{box}/(N_{\rm part.})^{1/3}$ leads to a slight delay in the onset
of reionization. Very small overdensities on scales below 0.5 Mpc/h,
that could harbor ionizing sources with HII regions around them, are
not captured by our analysis.

\section{Results for different time dependence of the ionization efficiency} 
\label{specific}

\begin{figure}
\bc
\epsfig{file=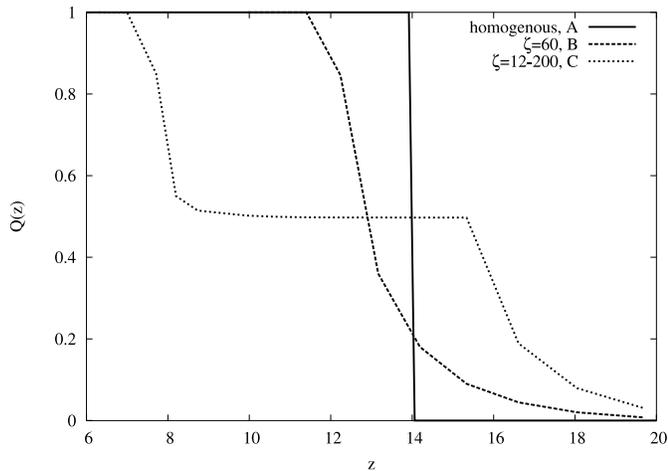,width=7cm,angle=-90}
\ec
\caption{Evolution of the fraction of ionized volume elements $Q(z)$
  in our calculations of reionization. Solid curve corresponds to
  instantaneous reionization at redshift $z=14$, yielding an optical
  depth of $\tau_{ri}\simeq 0.125$ for standard cosmological parameters. The
  dashed curve represents our patchy model in the case of a constant
  ionizing efficiency of sources. The dotted curve is for a model where a
  first generation of metal-free stars can survive the negative feedback of
  $H_2$ photodissociation for a limited period of time, then formation of
  these stars comes to a halt, leaving a network of HII regions. After a 
  while,
  Population II stars are born inside these regions and their photons gradually
  lead to a homogeneously ionized universe (Model C). Model C spends
  the longest time in the partly ionized regime and
  will therefore lead to the largest signal of the three models which
  have comparable total optical depth.}
\label{Qz}
\end{figure}

In our analysis we will compare three different reionization
scenarios. These are tuned such that they all lead to a comparable
integrated optical depth, $\tau_{ri} \simeq 0.125-0.133$. The
different histories of the fraction of ionized volume elements $Q(z)$
are shown in Figure \ref{Qz}.

\textbf{Model A } describes a universe that undergoes homogeneous 
and instantaneous 
  reionization at redshift 14. This is the standard scenario
  \citep{Seljak:1996is,Lewis:1999bs} assumed in likelihood analyses of
  large scale CMB polarization anisotropy (in its correlation with
  temperature, see \citealt{Kogut:2003et,Spergel:2003cb}). This model
  has $\tau_{ri}=0.133$.

\textbf{Model B } is the patchy model described above, with the
  assumption of constant source properties. An ionization efficiency
  $\zeta =60$ leads to reionization beginning at $z=19$ and concluding
  by $z=12$. This model has $\tau_{ri}=0.130$.

\textbf{Model C } exhibits extended patchiness. We describe this by 
assuming a first
  generation of metal free sources starting at redshift 20 with an
  ionization efficiency of $\zeta=200$ (it has been suggested that
  their photon output could be 10-20 times higher than that of normal
  stars, see
  \citealt{Schaerer:2002yr,Bromm:2001ag,Somerville:2003sk}). In our
  model, these sources sustain themselves only to redshifts around 15,
  because their hard photons dissociate fragile $H_2$ they need for
  cooling (e.g. \citealt{Mackey:2003au,Yoshida:2003au,Yoshida:2004au}).
  However, the first sources are assumed to leave pockets of
  ionized medium which begin to harbor Pop II stars. The initial mass
  function becomes less top heavy with time, and we describe the
  resulting Population II stars with an efficiency $\zeta=12$
  (see, e.g. Furlanetto et al. 2004c). This
  leads to a total optical depth of $\tau_{ri}=0.125$. Our model has a 
  monotonically evolving ionization fraction. Double
  reionization scenarios with an intermittent decrease of the
  ionization fraction have for instance been suggested in
  \citet{Cen:2003ey} and \citet{Wyithe:2003rr} but seem difficult to 
  realize in practice \citep{Furlanetto:2004nt}. We note that another way of
  reconciling CMB with quasar spectra observations is by resorting to
  recombinations that leave $0.1\%$ of the universe neutral until
  $z=6-7$
\citep{Sokasian:2003au,Sokasian:2004au}.

In Figure \ref{Qz}, we plot for all three models
the fraction of ionized volume elements as
a function of redshift, $Q(z)$. 
In Model A, the neutral fraction drops instantaneously at
$z=14$, in Model B sources of a constant ionization efficiency
$\zeta=60$ lead to a brief partly ionized phase, and in Model C the
ionization fraction ``freezes in'' for some time, while metal free
Pop. III stars cease to exist and leave residual ionized bubbles for
``normal'' Pop. II stars to be created and to ionize the medium fully
at $z=7-8$.

\begin{figure*}
\bc
\epsfig{file=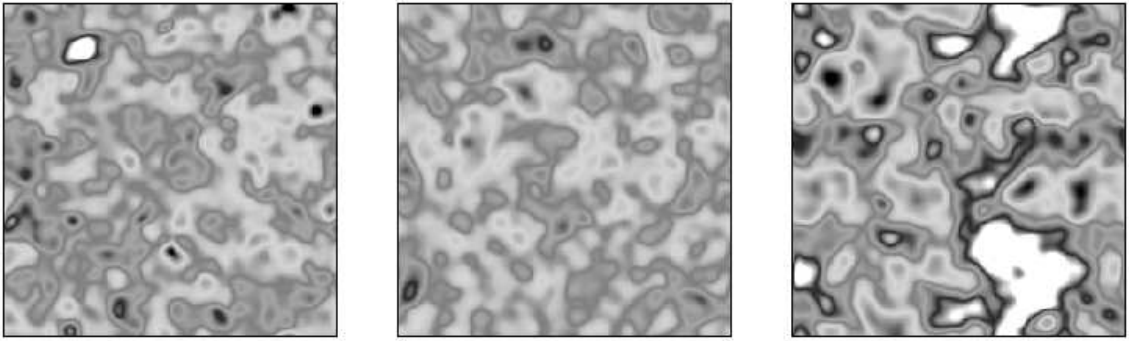,width=15cm}
\ec
\caption{Based on the implementation of Model B with a constant value of
  $\zeta$, we divide the Doppler signal into three redshift epochs.
  Plot on the left shows the kinetic Sunyaev-Zel'dovich effect out
  to redshift 3, in the middle the Ostriker-Vishniac regime from $z=3$
  out to $z=11$ is shown, and on the right the regime of non-uniform
  (patchy) reionization. Each picture shows the same angular extent,
  $1^\circ$. The left panel exhibits highly clustered
  structures. With increasing redshift, the clustering becomes more
  linear and patchy reionization leads to CMB signals on large
  scales. We smoothed the maps obtained from our simulations by a
  Gaussian beam corresponding to $l=10000\simeq 1'$, comparable to the
  angular resolutions of ACT and SPT.}
\label{3epochs}
\end{figure*}

\begin{figure}
\bc
\epsfig{file=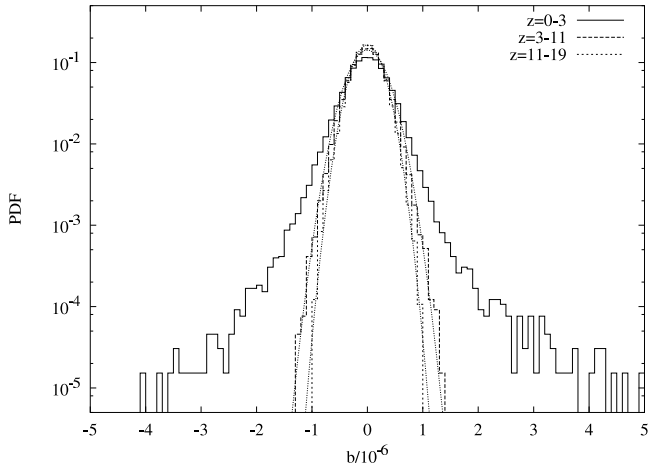,width=7cm,angle=-90}
\ec
\caption{The probability distribution functions for the three maps in Figure
  \ref{3epochs} are shown. The solid line is for the kinetic SZ regime
  z=0-3, the long-dashed line for the Ostriker-Vishniac era z=3-11,
  the short-dashed line for the patchy epoch of model B with constant
  ionization efficiency, z=11-19. The two histograms representing the
  earlier cosmological epochs are well-approximated by Gaussians.}
\label{3epochs-histo}
\end{figure}

We divide the total Doppler signal into three different redshift
regimes in Figure \ref{3epochs}. The maps have a side length of
$1^\circ$ and are smoothed with a Gaussian beam of width $\theta
\simeq 1'$, corresponding to a multipole number of $l_{max} \simeq
10000$, comparable to the angular resolution of ACT, and slightly
below that of SPT.  It is expected that during the interval from
$z=0$ to $z=3$, more strongly clustered regions make the Doppler effect
maps highly non-Gaussian. We show this by the solid line (PDF) in
Figure
\ref{3epochs-histo}. In the picture in the center ($z=3-11$)
structure formation is less advanced. The corresponding histogram can
be well-approximated by a Gaussian, as shown by the long-dashed
curve in Figure \ref{3epochs-histo}. Signals created by radially
moving inhomogeneities during this epoch are referred to as the
Ostriker-Vishniac effect \citep{OV}. The plot on the right of Figure \ref{3epochs} is the
patchy epoch of reionization Model B. Again, this epoch has a
largely Gaussian morphology (short-dashed line of Figure
\ref{3epochs-histo}), because structures on scales of the ionized
bubbles are just beginning to collapse. According to this figure, large
bubbles form on scales of tens of Mpc. The redshift regimes
represented by the left and middle picture have the same properties in
all three of our reionization scenarios.

\begin{figure}
\bc
\vspace{-1cm}
\epsfig{file=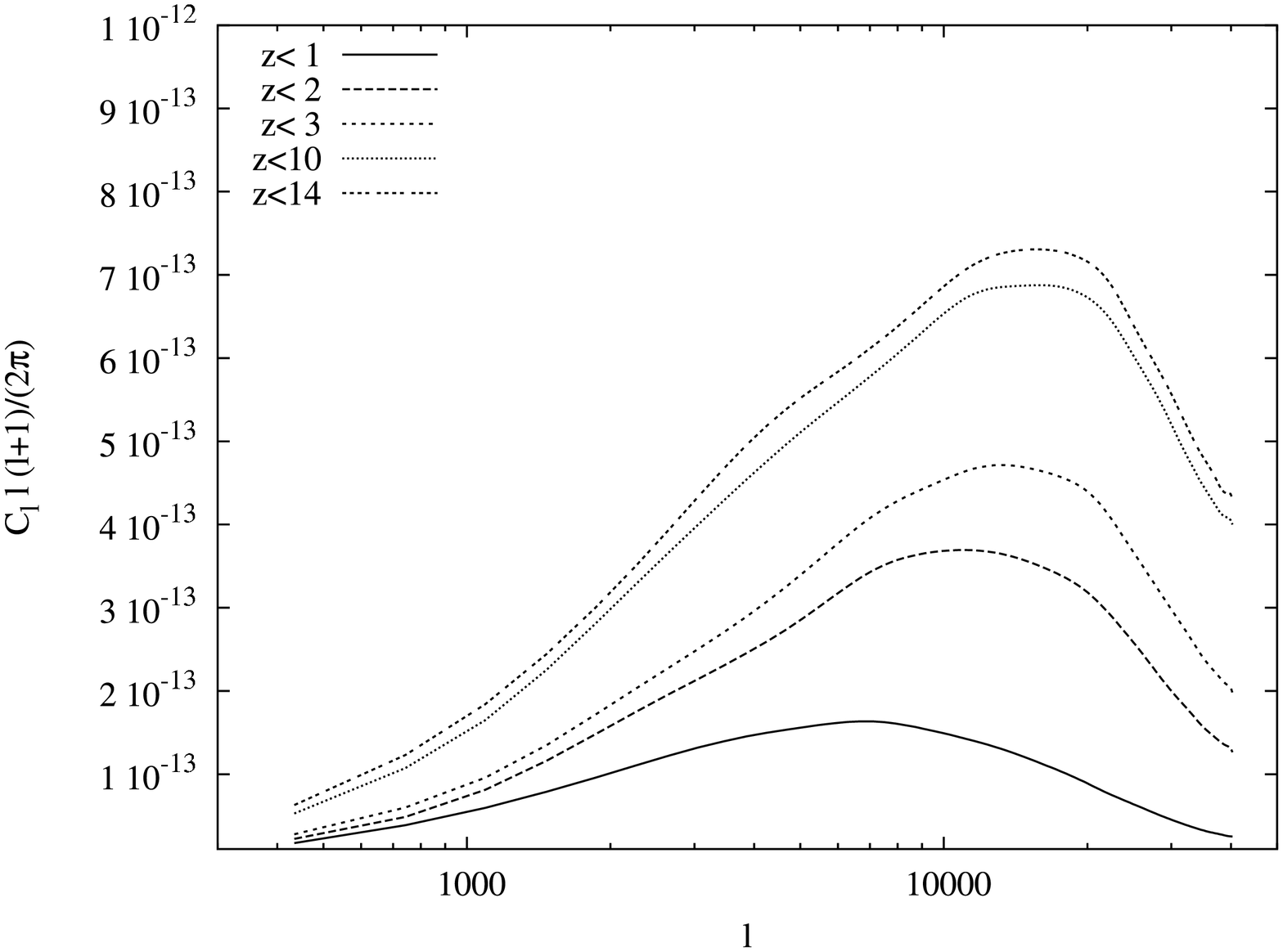,width=7cm,angle=-90}
\epsfig{file=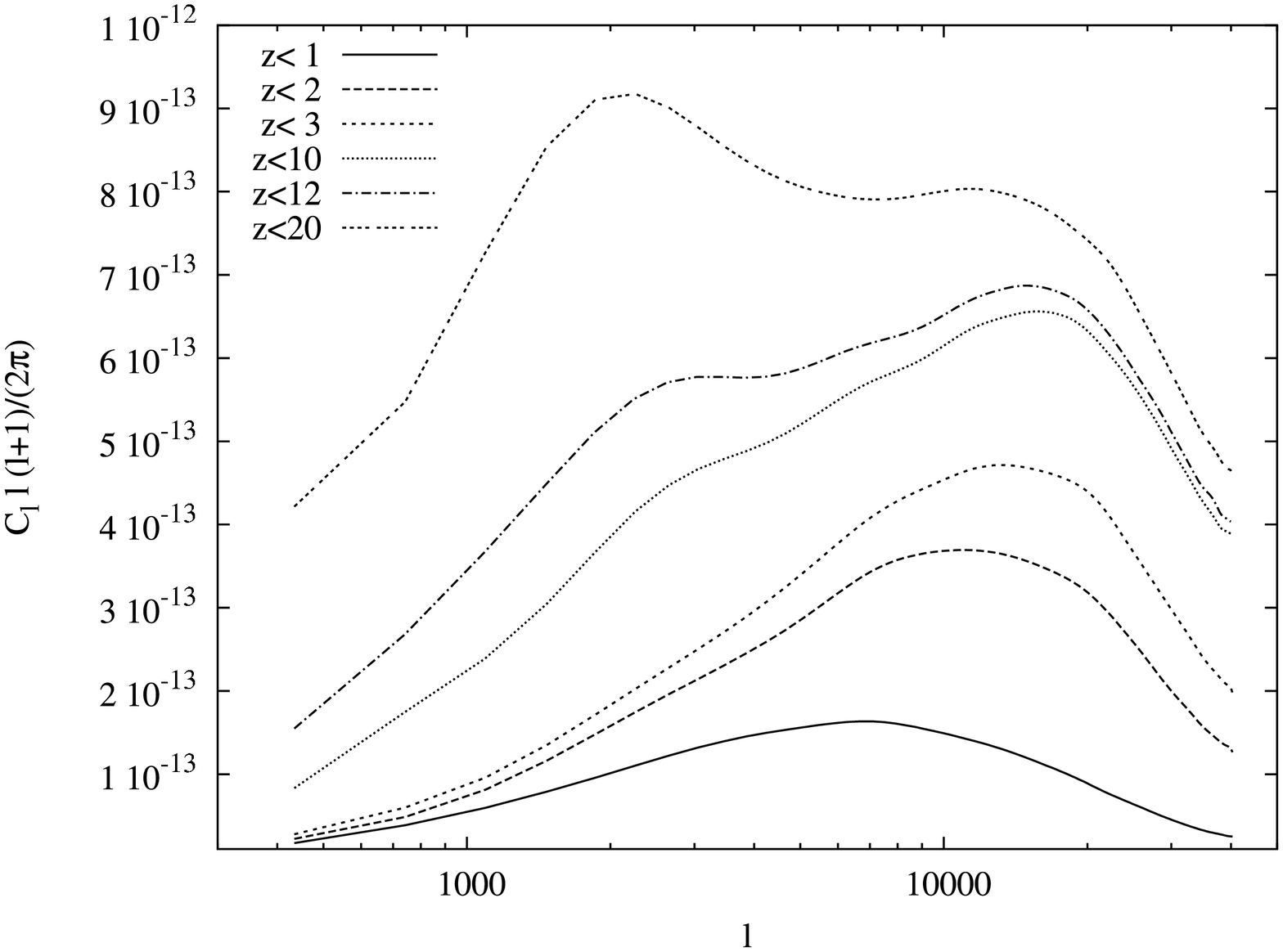,width=7cm,angle=-90}
\ec
\caption{Contribution to the kinetic SZ signal for Models
  A (upper panel) and C (lower panel) out to different redshifts.
  Although a large
  fraction of the total Doppler effect comes from low redshift, high
  overdensity regions, there is a significant contribution out to
  $z\simeq 10$. In the patchy model (bottom) the regime
  $z=11-19$ leads to further enhancement of the signal.}
\label{z_depend}
\end{figure}

The upper panel of Figure \ref{z_depend} shows that in the
homogeneously reionizing universe, the largest portion of the Doppler
signal comes from low redshifts, the kSZ regime. The upper panel
represents the case of homogeneous reionization, Model A. Contrary to
the thermal SZ effect,
the signal still carries information about higher
redshifts, $z \leq 10$. We can use this principle to gain knowledge about
the details of reionization taking place at high redshifts. The
additional contributions from larger redshifts ($z > 10$) in Model
C are shown in the lower panel of the figure.

\begin{figure}
\bc
\epsfig{file=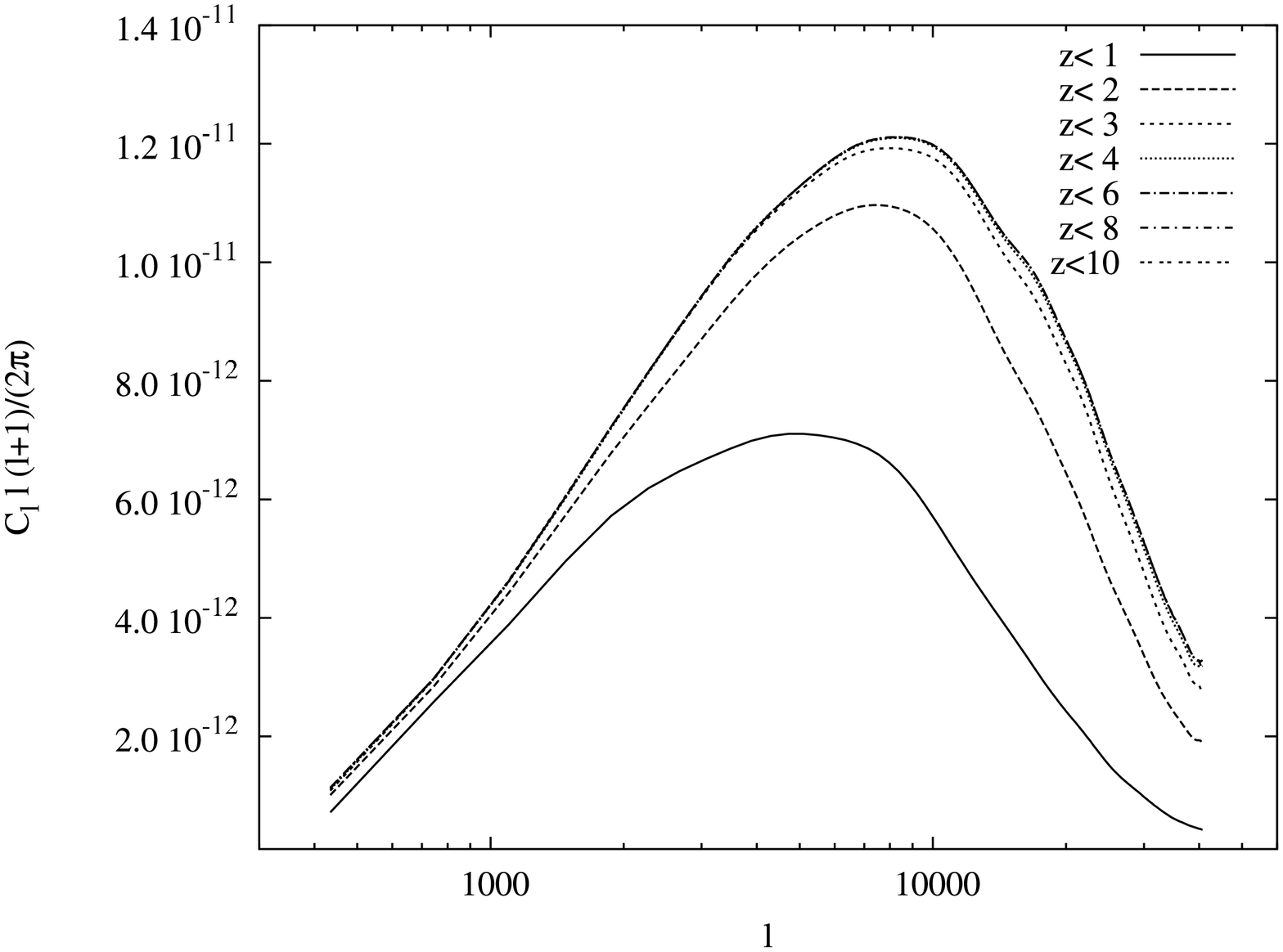,width=7cm,angle=-90}
\ec
\caption{Redshift dependence of the thermal
  Sunyaev-Zel'dovich $\Delta T/T =-2 y$ effect as generated with our
  simulations. The major contribution comes from galaxy clusters in
  the regime $z<3$. Hence, the thermal SZ effect
  is less suited for studying the reionization epoch than is the kinetic SZ
  effect.}
\label{y_zdepend}
\end{figure}

For comparison, we show in Figure \ref{y_zdepend} how different
redshift regimes contribute to the thermal SZ effect (where $\Delta
T/T=-2 y$ for low frequencies). The plot shows that this signal is
basically saturated at $z=3$; the thermal SZ effect receives most of
its contribution from galaxy clusters in the more nearby universe.

\section{Observability of patchy reionization with future CMB 
experiments}
\label{exper}

In this section, we assess the observability of patchy
reionization with future experiments. ACT and SPT should be able to
distinguish different reionization scenarios with high significance by
measuring temperature power spectra. In their cosmological parameter
analyses, all-sky experiments such as Planck will have to 
account for patchy
reionization as a possible source of bias and need to
rely on their polarization data in order to obtain proper
constraints.

\subsection{Power spectral constraints from ACT and SPT}
\label{power}

\begin{figure*}
\bc
\epsfig{file=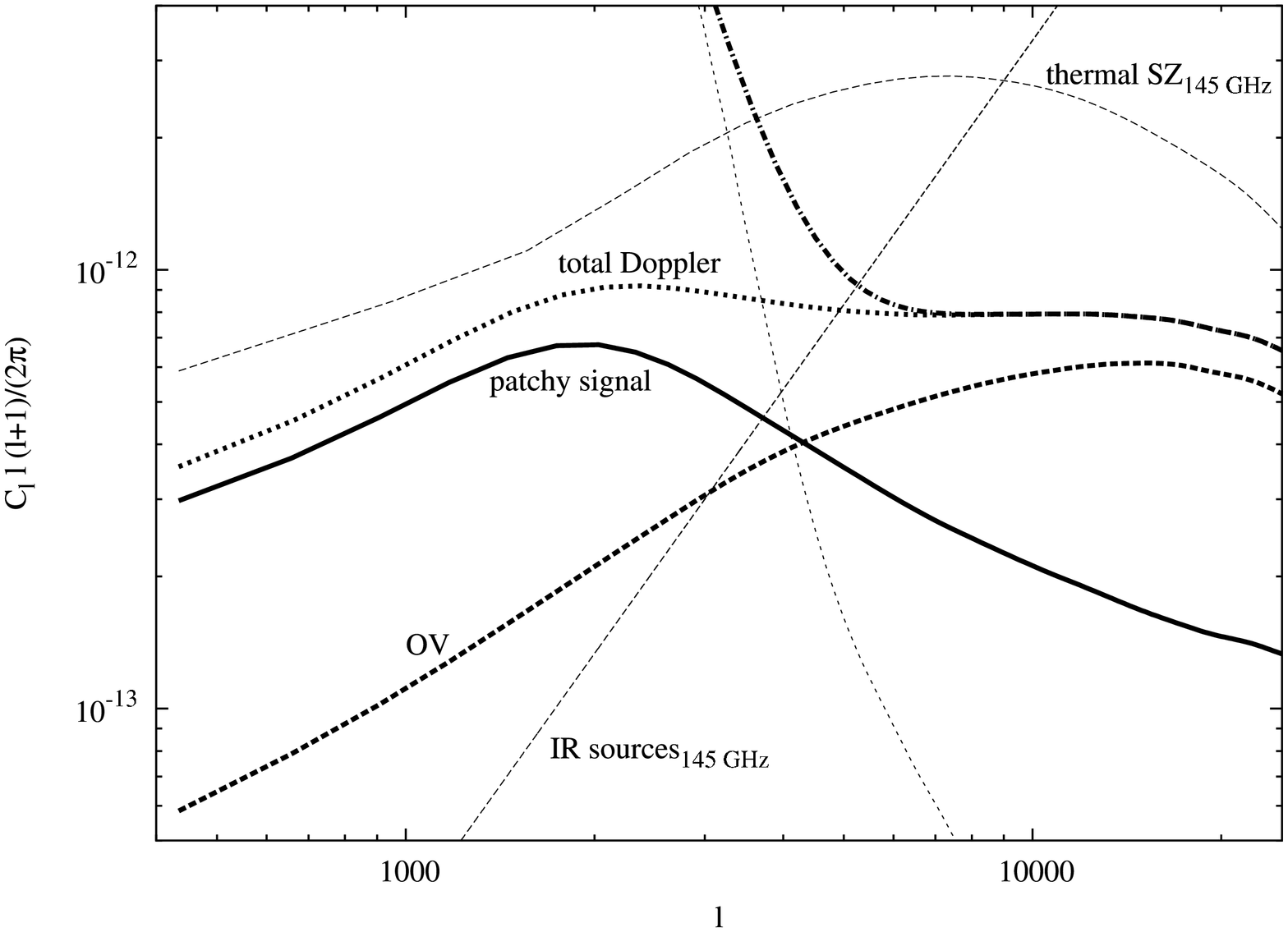,width=10.0cm,angle=-90}
\ec
\caption{Different contributions to the total Doppler
  signal (kinetic SZ) in our extended patchy reionization Model C. The
  solid curve is the contribution to temperature anisotropies from the
  patchy regime alone at $z=7-20$. The dashed curve gives the Doppler
  effect from density modulations at homogeneous ionization out to
  $z=11$. The dotted curve sums up those contributions to the
  total Doppler signal. The total temperature fluctuations, cleaned of
  the thermal SZ and IR sources (shown at 145 GHz in the thin dashed lines), is given
  by the dot-dashed curve. These curves are smoothed versions of the power
  spectra generated from 50 maps.}
\label{Cl_contribs}
\end{figure*}

The effect of patchy reionization on the CMB power spectrum is of
similar magnitude to that of the Doppler effect induced by
variations in the mean density. The first source population is heavily
biased so the patchy reionization signal peaks on larger angular
scales.  Figure \ref{movie} shows that for the regime in which the
ionization fraction is roughly $50\%$ (which is where most of the
signal comes from) the bubbles reach sizes of tens of comoving Mpc.
In Figure \ref{Cl_contribs} we plot the different contributions to the
Doppler power spectrum. Patchy reionization in Model C with
$\zeta=12-200$ lasts from redshifts 20 to 7. The signal imprinted in
the CMB from this epoch is shown in solid.  At later times, the
universe is homogeneously reionized and the dashed line shows the
kinetic SZ effect for this period.

Inverse Compton scattering in galaxy clusters leads to a larger
signal than that caused by Doppler scattering. Because only a small
fraction of the CMB photons present are likely to scatter inside the
cluster medium ($\simeq 1\%$ for a massive 15 keV cluster), the
distribution will not thermalize to that of the hot gas. The effect has a
characteristic frequency dependence, and it can be distinguished from
Doppler scatterings which leave the frequency distribution of the
photons unchanged.  The dot-dashed line in Figure \ref{Cl_contribs} is the
combination of the secondary signal with the primordial CMB, once the
thermal SZ with its characteristic frequency distribution has been
removed.

On scales where patchiness during reionization contributes more to the
total signal than the Doppler effect owing to modulations in the
density, ($l < 4000$), the primordial anisotropies dominate. To study
the secondary anisotropies created during reionization one has to
observe angular scales where photon diffusion smoothes out the primary
anisotropies.

Upcoming experiments such as ACT and SPT will observe the sky in a
number of frequency bands with comparable sensitivity and angular
resolution.  ACT \citep{Kosowsky:2004sw} will observe in three frequency bands: 145, 225, and
265 GHz, with FWHM beam-widths $\theta_{FWHM}$ of 1.7, 1.1, and 0.9
arcminutes, respectively. We use the specifications of the 225 GHz
channel for our power spectrum analysis because the thermal SZ effect
almost vanishes in this frequency regime, its zero being at $\nu
\simeq 218\ GHz$. The sensitivity per resolution pixel for this
channel is $\sigma=2 \ \mu K$.  The South Pole Telescope \citep{Ruhl:2004kv} will observe
in five bands at 95, 150, 219, 274, and 345 GHz. The 219 GHz channel
will have an angular resolution of $\theta_{FWHM}=0.69'$ and a
sensitivity of $10 \mu K$ \footnote{J.E. Ruhl,
private communication.}. We assume sky coverages of $0.5\%$ and
$10\%$ for ACT and SPT, respectively.

From these specifications and a template for the power spectrum of
primary and secondary anisotropies at arcminute scales (which we
obtain from our simulations), we can calculate the errors in the $C_l$
determination, including noise as additional random field on the
sky \citep{Tegmark:1997vs}):
\beq
\Delta C_l = \sqrt{\frac{2}{f_{sky} (2l+1)}} \left[ C_l + \frac{f_{sky}}{w B_l^2} \right],
\label{Cl_error}
\eeq
where we assume errors bars corresponding to a Gaussian map. 
Here, $w=(\theta_{fwhm} \sigma)^{-2}$, and $B_l$ is a beam
profile (assumed Gaussian) given by $B_l=e^{-\theta_b^2 l (l+1)/2}$
(with $\theta_b=\theta_{fwhm}/\sqrt{8 \ln 2}$).

We assume that experiments such as ALMA
\footnote{http://www.eso.org/projects/alma/science/alma-science.pdf}
will lead to a good understanding of the point source frequency spectrum and angular clustering.
In more pessimistic scenarios where only the
frequency dependence or nothing is known about point sources, they can
strongly degrade our ability to detect the kinetic SZ
\citep{Huffenberger:2004gm} (we included their estimate for the IR source power spectrum at 145 GHz in Figure \ref{Cl_contribs}). 
On the other hand the assumption of perfect cleaning of the thermal SZ effect is safer because we understand its frequency dependence well.

\begin{figure*}
\bc
\epsfig{file=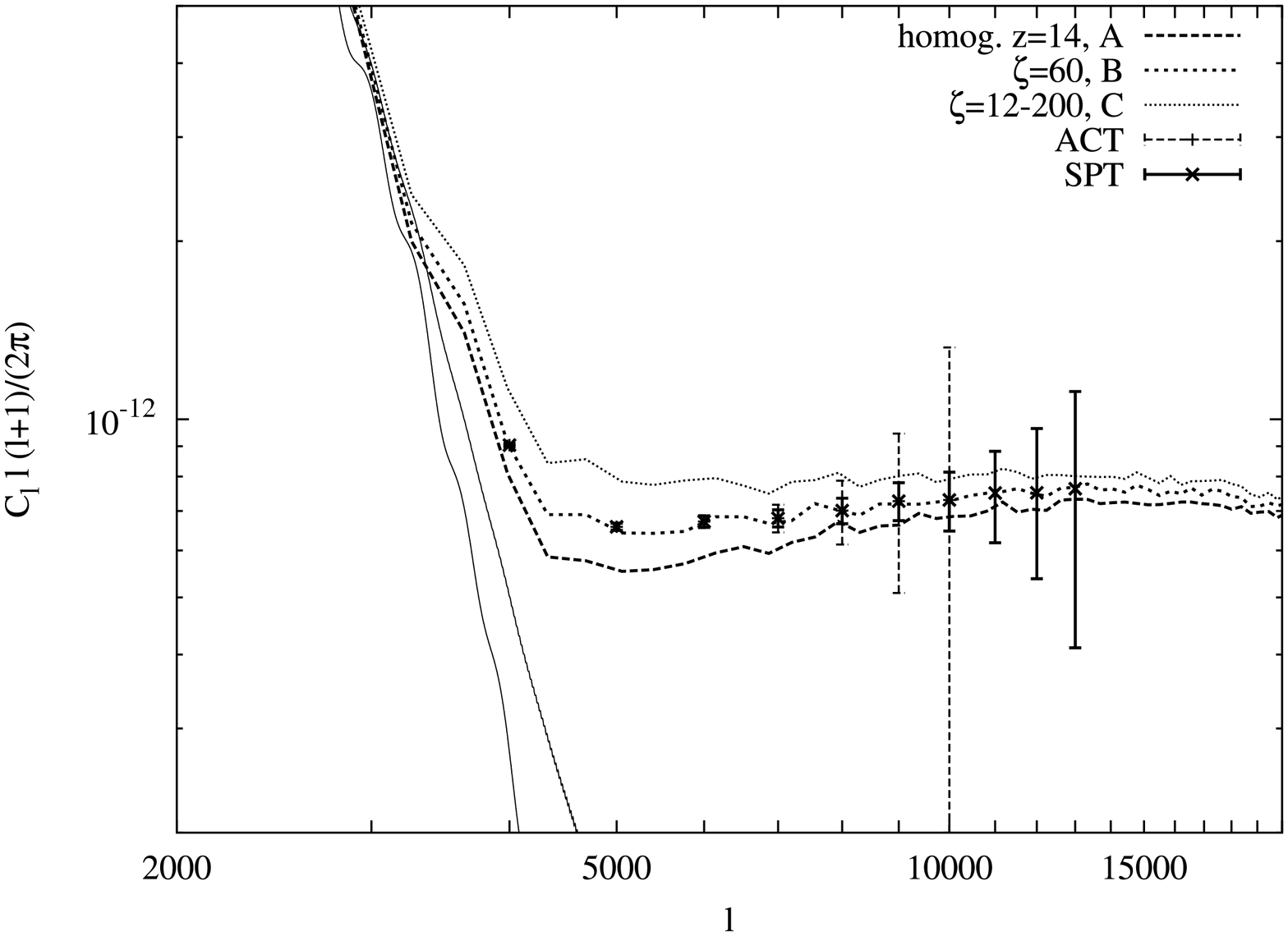,width=10.0cm,angle=-90}
\ec
\caption{With future experiments like ACT, the various reionization
  models should be easily distinguishable by measuring the temperature
  power spectrum. Since the kinetic SZ signal is almost featureless on
  the scales of experimental relevance, the band-powers can be
  combined to a simple amplitude when distinguishing the different
  models. We find that the double reionization scenario (given by the
  upper power spectrum) could thus be distinguished at high significance
  from a uniform, instantaneous model where the universe reionizes at
  $z=14$ (bottom curve). The change of the ionizing efficiency with
  time $\dot{\zeta}$ affects the slope of the secondary anisotropy
  power spectra.}
\label{distinguish-pow}
\end{figure*}

To compare the experimental constraints with the power spectra
extracted from our simulations on degree patches on the sky, we bin the errors into bands of width $\Delta l=360$. The predicted
measurements errors are plotted in Figure
\ref{distinguish-pow}, together with power spectra generated from the
``cleaned'' maps of primordial CMB and kinetic SZ combined.  Since the
kinetic SZ appears almost featureless on the scales accessible to the
next generation of experiments, reionization models can be
distinguished by their average amplitude on these scales.  The error
bars in the relevant band-powers of ACT are combined using
$\sigma^2_{tot}=({\sum_i \sigma^{-2}_i})^{-1}$. It follows that the
overall amplitude can be measured with an accuracy of $\sigma_{\Delta
T}=0.011 \mu K$ (we assumed a $1\%$ calibration uncertainty). Given
that the plateau of the extended patchy model (Model C) lies at
$\Delta T \simeq 2.407 \mu K$, while Model A has an average amplitude
on these scales of $\Delta T \simeq 2.076 \mu K$, the two will be
distinguishable at the $30 \sigma$ level with ACT, if we ignore contamination by point 
sources and by the thermal SZ, which is not a realistic assumption. This is the same
for the South Pole Telescope, which has better angular resolution, but will look
less deep (it has a much larger survey area).

It is also of interest to ask whether additional constraints on reionization
scenarios could be achieved by using non-Gaussian statistics, in particular 
the four point function (the skewness arising from a
line-of-sight velocity effect should vanish). It may be expected, that
because the HII regions are created inside large overdensities at high
redshift, patchy models are more Gaussian on the whole. The main
problem with this notion is that in any patchy model with a comoving bubble size comparable to ours this effect will suffer from ``washing out'' by the primordial CMB. On scales where Doppler induced
fluctuations become larger than the primordial CMB, the
impact of patchiness amounts only to a fraction of the total signal
(15\% in Model B, 30\% in Model C), the rest being attributed to
scatterings owing to density modulations alone. Also, the majority of non Gaussian contributions comes from clusters and filaments at $z < 3$, compare Figure \ref{3epochs-histo}.
We computed the kurtosis
\beq
\Theta_4 \equiv \frac{\langle (\Delta T/T)^4 \rangle}{\sigma_T^4} - 3
\eeq
of thermal SZ ``cleaned'' maps (i.e. primordial CMB+Doppler) that were 
filtered with a Gaussian in Fourier space, approximately cutting out
``contamination'' by the primordial CMB anisotropies below $l\simeq
3500$, and beam smearing at scales of $l\simeq 9000$ comparable to the
angular resolution of ACT (similar to the window function proposed by
\citealt{Huffenberger:2004gm}). 
Using this method, we do not find a statistically significant difference 
between patchy and homogeneous models of reionization.

\subsection{Expected bias in cosmological parameter determination from
  Planck}
\label{fisher}

Cosmic microwave background measurements have been and will likely
remain the most precise tools for the measurement of
cosmological parameters. The best constraints will come from a
combination of CMB temperature and polarization power spectra which
encapsulate all the relevant information in the sky maps. The aim is
to make the data cosmic variance limited to as high as possible
multipole numbers. WMAP is cosmic variance limited up to $l \simeq
500$.

The Planck satellite should achieve cosmic variance limitation out to
$l=2500$ for its temperature power spectrum measurement.  Hence,
Planck will reach into the regime where the secondary anisotropies
become important. To avoid biases in parameter estimation, systematic
changes that the secondaries may produce in the power spectra need to
be considered.

The contribution to the power spectrum from patchy reionization is one
to two orders of magnitude smaller than the primordial CMB
anisotropies on the relevant scales, but owing to the exquisite sensitivity
of these experiments it biases parameter estimates.
For the analytic models of patchy reionization suggested by
\citet{Knox:1998fp} and \citet{Santos:2003jb} this parameter bias was estimated.  In these models the mean bubble size is smaller than in our computation, and the signal peaks at higher multipoles. 

The bias can be estimated from the Fisher matrix coefficients
\beq
F_{ij}=\sum_l \frac{\partial C_l}{\partial p_i} w_l
\frac{\partial C_l}{\partial p_j}
\eeq
in the following manner:
\beq
\mathcal{B}_i \equiv \frac{\Delta p_i}{\sigma_i}=\frac{\sum_j F_{ij}^{-1} \sum_l
  \frac{\partial C_l}{\partial p_j} C_l^{\rm D}}{\sqrt{F_{ii}^{-1}}}
\eeq
where $\Delta p_i$ are the systematic biases in the determination of
parameters $p_i$, $w_l$ are the inverse squares of the statistical
errors in the $C_l$ estimation, given by Equation
\ref{Cl_error} and ${\sigma_i}=\sqrt{F_{ii}^{-1}}$ is the estimate 
of the error bars for parameter $i$. 
In this expression, $C_l^{\rm D}$ denotes the combined
Doppler power spectrum owing to kSZ/OV and patchy reionization.

In Table \ref{bias} we show the results of our analysis,
using the power spectra for our different models as bias \footnote{We combined
the three frequency channels of Planck with the highest angular resolution and
took into account the number of polarized instruments. For a one year 
observation period, the three
channels (217, 143 and 100 GHz) have $\theta_{{\rm fwhm}}=$ 5.0,7.1 and
9.2 arcminutes. This leads to the raw sensitivities:
\begin{eqnarray*}
w_T^{-1} &=& (0.0084 \mu K)^2 \\
w_P^{-1} &=& (0.0200 \mu K)^2 \, .
\end{eqnarray*}
We assumed a sky coverage of $f_{Sky}=0.8$.}. 
The first line shows the statistical error alone,
$\sqrt{F_{ii}^{-1}}$, the following lines show the parameter bias
$\mathcal{B}_i$ for Models A, B, and C. For Planck, we used the
specifications of the High Frequency Instrument
\footnote{http://www.rssd.esa.int/index.php?project=PLANCK\&page=perf\_top}.
The
power spectra derivatives for the Fisher analysis were computed for
the fiducial cosmological model given in the introduction.

\begin{table}
\bc
\begin{tabular}{|c|c|c|c|c|c|c|}
\hline
\bf{model} & $\tau_{ri}$ & $\Omega_\Lambda$ & $\omega_{dm}$ & $\omega_{b}$ & $n_s$ &
$A_s$ \\
\hline
$1 \sigma$ & 0.0035 & 0.010 & 0.0017 & 0.00018 & 0.0045 & 0.0050 \\
\hline
 A & 0.40 & 1.18 & -1.10 & 1.71 & 2.14 & 0.16 \\
 B & 1.26 & 1.66 & -1.54 & 2.40 & 3.05 & 0.78 \\
 C & 2.25 & 2.89 & -2.69 & 4.20 & 5.62 & 1.58 \\
\hline
\end{tabular}
\ec
\caption{Bias in units of the statistical error ($\mathcal{B}_i$) 
expected for cosmological parameter estimation
  with Planck, if
  temperature and polarization power spectra are used and the
  influences of kSZ/OV and
  patchy reionization are neglected in the power spectrum analysis. The
  maximum multipole in our analysis was $l=4000$.}
\label{bias}
\end{table}

It is clear from Table \ref{bias} that secondary anisotropies need to
be taken into account.  This is the case in particular for parameters
that influence the shape of the power spectra at intermediate scales
($\omega_{b},\omega_{dm},n_s$). Constraints on the amplitude $A_s$ and
the optical depth owing to reionization $\tau_{\rm ri}$ are less
heavily biased (besides the amplitude, the reionization optical depth
only affects CMB polarization on large scales). Note that even in a
simple homogeneous reionization model, as our Model A, most of the
biases are of order unity. 
The individual systematic shifts depend strongly on what parameters are used. 
The analysis of \citet{Santos:2003jb} has extra parameters that make the biasing source (the patchy power spectrum) distribute differently into each parameter offset.

The polarization anisotropies generated during the reionization epoch
are expected to be four orders of magnitude smaller than the
temperature power spectra \citep{Hu:1999vq}.  This suggests that to
avoid biases one could use temperature information down to an angular
scale where the Doppler contamination is still negligible but use the
polarization data for all $l$. We show the results of an analysis in
which we used temperature information only out to $l\simeq 1000$, but
polarization in the full range accessible to Planck (this will be out
to $l\simeq 1800$) in Table
\ref{bias_reduced}. The $1 \sigma$ error bars for the parameters are
only slightly larger than in Table \ref{bias}, indicating that the
parameter estimates are more or less ``saturated'' at $l\simeq
1000$. On the other hand, the expected parameter biases owing to
reionization are much smaller. The use of polarization information
in future CMB experiments thus can play an important role beyond breaking
degeneracies between traditional cosmological parameters and improving
the error bars.

\begin{table}
\bc
\begin{tabular}{|c|c|c|c|c|c|c|}
\hline
\bf{model} & $\tau_{ri}$ & $\Omega_\Lambda$ & $\omega_{dm}$ & $\omega_{b}$ & $n_s$ &
$A_s$ \\
\hline
$1 \sigma$ & 0.0041 & 0.013 & 0.0020 & 0.00026 & 0.0079 & 0.0056 \\
\hline
A & 0.010 & -0.018 & 0.017 & 0.024 & 0.086 & 0.068 \\
B & 0.027 & -0.028 & 0.026 & 0.037 & 0.135 & 0.196 \\
C & 0.050 & -0.070 & 0.067 & 0.058 & 0.301 & 0.487 \\
\hline
\end{tabular}
\ec
\caption{Bias in cosmological parameter estimation
  with Planck is reduced significantly, when
  temperature power spectra are used only until $l=1000$ and
  polarization power spectra are used in the whole range (Planck
  should be able to measure $C_l^{EE}$ out to $l\simeq 1800$). 
  On the other hand,
  the parameter errors become only slightly larger by leaving out the
  high $l$ temperature information.}
\label{bias_reduced}
\end{table}

The other strategy is to include an extra parameter to model the
effect of the Doppler contributions. This approach has the added
advantage that a positive constraint on reionization could be obtained
with Planck alone.  Patchiness shifts the power spectrum on small
scales, and we model this by a ``patchy amplitude'' parameter. This
parameter is included by adding the Doppler spectra given in the
last section with a variable amplitude $A_D$ to the primordial CMB. If
the parameter derivatives in the Fisher matrix center ($A_D=1$) around
Model B, we find that the amplitude parameter could be constrained
with $\sigma_{A_D} =0.51$. The model is only $\simeq 0.3 \sigma$ away
from homogeneous reionization (Model A), so Planck will not be able to
make a strong statement. If patchy reionization turns out to be
extended, similar to Model C, Planck should observe this at higher
significance $\sigma_{A_D}=0.20$, amounting to a $5\sigma$ detection
of Doppler induced secondary anisotropies, with the amplitude of Model
C being $3 \sigma$ away from homogeneous reionization. Finally, if
reionization proceeded homogeneously, Planck will not gain knowledge
beyond its large scale polarization measurements of $\tau_{ri}$ by
using small scale temperature fluctuations, given that the uncertainty
for our Model A lies at $\sigma_{A_D}=0.71$.

When Planck's power spectra are modeled with the ``patchy amplitude''
parameter, constraints for the standard cosmological parameters are
slightly improved over the analysis that abandoned temperature data
beyond $l=1000$ (Table \ref{bias_reduced}), despite the introduction of
an extra parameter. Concretely, we find in this analysis that
$\sigma_{\tau_{ri}}$=0.037, $\sigma_{\Omega_\Lambda}$=0.011,
$\sigma_{\omega_{dm}}$=0.0018, $\sigma_{\omega_b}$=0.00021,
$\sigma_{\Omega_{n_s}}$=0.0058, $\sigma_{A_s}$=0.0052.

The Doppler power spectrum can bias the result of parameter analyses
with future CMB surveys. Planck may be able to improve constraints on
models with an extended epoch of patchiness.  We hope to have shown in
this section that careful modeling of the epoch of first stars will
become crucial for doing precision cosmology with progressively more
refined CMB experiments.

\section{Discussion}

We have presented simulations of secondary anisotropies in the cosmic
microwave background calculated using smoothed particle hydrodynamics
simulations of large scale structure, focusing on the effect
produced by patchy reionization. We incorporated an analytic model for
the morphology of the HII regions into our numerical treatment and
investigated whether such patchy scenarios can be distinguished from
homogeneous reionization.

An important advantage of our technique over pure analytical
predictions of patchy reionization morphology is that we follow
the complicated clustering of dark matter and baryons into the
slightly non-linear regime.
In contrast to full radiative transfer calculations of the reionization
epoch, we combine uncertainties in the physics of source
formation, feedback processes and radiative transfer 
into a single parameter and explore the consequences of varying this
parameter.  This simplification allows us to make predictions on
scales an order of magnitude larger than current radiative transfer
schemes can accomplish with a small expense of memory and CPU.

We extracted power spectra from sky maps produced by tracing rays
across our simulation volumes. The patchy reionization signal peaks on
multipole scales of $l\simeq 2000$, and it increases the amplitude of the ``cleaned'' CMB power spectrum by up to $30\%$ on scales $l\geq4000$, so that the total
level of Doppler related anisotropies is $\Delta T \simeq 2.4 \mu
K$. We found that with the next generation of ground based CMB
experiments (ACT, SPT) the different reionization models we
investigated could be distinguished with high significance by using
the power spectrum. Additional information about the morphological
properties of that epoch that could be obtained for instance by
measuring the four point function or other deviations from Gaussianity
will probably be difficult to obtain, because the reionization signal
peaks on angular scales where the primordial CMB anisotropies
dominate.  In the future, it may be possible to combine measurements
of the CMB with other observations such as 21 cm fluctuations from
neutral hydrogen \citep{Zaldarriaga:2004cm,coo04}
or Lyman-alpha emission from high redshift galaxies (e.g. \citealt{Furlanetto:2004to}) to further constrain the topology of the reionization
process.

We investigated the bias in the determination of cosmological
parameters that will be produced by the additional patchy reionization
signal, when extracting cosmological parameters from CMB anisotropy
measurements. We find that for Planck this bias is significant.  The
bias may be circumvented by focusing completely on polarization
information in the multipole regime where patchiness peaks, with only
a slight disadvantage in parameter constraints. Alternatively, a
template for the Doppler spectrum could be introduced in the parameter
analysis which may lead to a detection of the effects of an extended
reionization phase.

\acknowledgments
O. Z. and M. Z. are supported by NSF grant AST 0098606 and by the David and Lucille Packard Foundation Fellowship for Science and Engineering and by the Sloan Foundation. 
This work is also supported by NSF grants ACI 96-19019, AST
00-71019, AST 02-06299, and AST 03-07690, and NASA ATP grants
NAG5-12140, NAG5-13292, and NAG5-13381.  The simulations were
performed at the Center for Parallel Astrophysical Computing at the
Harvard-Smithsonian Center for Astrophysics.

\end{document}